\documentclass[prl, superscriptaddress, twocolumn]{revtex4-2}

\usepackage{fullpage}
\usepackage{setspace}
\usepackage{parskip}
\usepackage{titlesec}
\usepackage{xcolor,colortbl}
\usepackage{lineno}
\usepackage[colorlinks = true,
            linkcolor = blue,
            urlcolor  = blue,
            citecolor = blue,
            anchorcolor = blue]{hyperref}
\usepackage{etoolbox}

\titlespacing{\section}{0pt}{*3}{*1}
\titlespacing{\subsection}{0pt}{*2}{*0.5}
\titlespacing{\subsubsection}{0pt}{*1.5}{0pt}

\usepackage[caption=false]{subfig}
\usepackage{graphicx}
\usepackage{latexsym}
\usepackage{booktabs,array,multirow}
\usepackage{amsfonts,amsmath,amssymb}
\usepackage{cuted}

\usepackage{mathtools,nccmath}

\providecommand\citet{\cite}
\providecommand\citep{\cite}

\usepackage[utf8]{inputenc}
\usepackage[english]{babel}

\newcommand{\ra}[1]{\renewcommand{\arraystretch}{#1}}

\usepackage{tabularx}

\definecolor{Gray}{gray}{0.9}
\definecolor{LightCyan}{rgb}{0.88,1,1}
\definecolor{LightBlue}{rgb}{0.5,0.5,0.8}

\makeatletter
\newcommand{\colorcaption}[2][]{%
  \begingroup%
  \renewcommand{\@caption@fignum@sep}{ (color). }%
  \caption[#1]{#2}%
  \endgroup%
}
\makeatother

\usepackage{appendix}

\begin{document}

\setlength{\abovedisplayskip}{3pt}
\setlength{\belowdisplayskip}{3pt}

\def\figurename{FIG.}
%\def\@caption@fignum@sep{ (color). }

% \title{Microscopic force transmission due to hydrogel on a rigid substrate}
\title{A density functional theory approach to interpret elastowetting of hydrogels}
\author{Priyam Chakraborty,$^{1,\ast}$ Surjyasish Mitra,$^{1,\ast}$ A-Reum Kim,$^{2,\ast}$, Boxin Zhao,$^{2,\ast\ast}$ and Sushanta K. Mitra $^{1,\ast\ast}$\\
\vspace{0.5 cm}
\normalsize{$^{1}$ Micro \& Nano-scale Transport Laboratory,
Department of Mechanical and Mechantronics Engineering,}
\normalsize{ Waterloo Institute for Nanotechnology, University of Waterloo,}
\normalsize {200 University Avenue West, Waterloo, ON N2L 3G1, Canada}\\
\normalsize{$^{2}$Department of Chemical Engineering, Waterloo Institute for Nanotechnology, University of Waterloo,}
\normalsize {200 University Avenue West, Waterloo, ON N2L 3G1, Canada}\\
\vspace{0.5 cm}
\normalsize{$^\ast$These authors contributed equally to this work}\\
\normalsize{$^{\ast\ast}$Corresponding author: zhaob@uwaterloo.ca, skmitra@uwaterloo.ca}}

\begin{abstract}
Sessile hydrogel drops on rigid surfaces exhibit wetting/contact morphology intermediate between liquid drops and glass spheres. 
Using \emph{density functional theory}, we reveal the contact forces acting
between a hydrogel and a rigid glass surface. We show that while transitioning from liquid-like to solid-like hydrogels, there exists a critical 
hydrogel elasticity which enables a switch from attractive to repulsive  interaction with the underlying rigid glass surface. Our theoretical model is validated by experimental observations of sessile Polyacrylamide (PAAm) hydrogels of varying elasticity on glass surfaces. Further, the proposed model successfully approaches Young's law in the \emph{pure} liquid limit and work of adhesion in the \emph{glassy} limit. Lastly, we show a modified contact angle relation taking into account the hydrogel elasticity to explain the features of a distinct \emph{hydrogel foot}.
 
\end{abstract}%

\maketitle

% \section{Introduction}

% What is the significance of studying forces at the contact between droplets and surfaces? 

% How does the physics differ if the droplet is a hydrogel instead of a pure liquid? 

% What are the open areas of research in force transmission between hydrogel and solid surface? 
\section{I. Introduction}
Intermediate between solids and liquids, hydrogels offer unique properties in terms of their chemistry as well as tunable elasticity \cite{yang2020hydrogel,sun2012highly,gao2019photodetachable}. 
Consequently, study of hydrogels has been at the forefront
of emerging technologies like flexible electronics \cite{le2017wearable,yuk2016skin,wei2021multifunctional}, advanced manufacturing  \cite{liu2018bonding}, 3D printing \cite{bauman2023multi}, and biomedical devices \cite{kim2010dynamic,zhang2020catechol}. 
However, majority of existing literature have focused on characterizing material properties and surface chemistry of hydrogels \cite{yang2020hydrogel,sun2012highly,kim2010dynamic}. 
In many practical and engineering applications such as cell mechanics \cite{schwarz2013physics,style2013patterning}, understanding how hydrogels wet any surface is crucial and remains largely unexplored \cite{chakrabarti2018elastowetting}.
Owing to their finite viscoelasticity, wetting of hydrogels
is expected to deviate significantly from the equilibrium
picture of Newtonian liquid drops.
The classical Young's law describes the equilibrium
configuration of a liquid drop on a rigid substrate \cite{young1805iii}.
Varying the substrate elasticity, i.e., using soft substrates, reveals a markedly different
equilibrium configuration where liquid capillarity and substrate elasticity competes at the three phase contact line forming the \emph{wetting ridge} profile \cite{marchand2012contact,jerison2011deformation,
style2013universal,pericet2009solid,pericet2008effect,mitra2022probing}. Here, use of hydrogels provides us
with a mechanism to vary droplet elasticity instead. 

In this work, we have a unique proposition where we reveal 
the competition of elasticity and capillarity occuring
within the hydrogel (drop) phase, rather than the widely studied liquid drops in contact with soft substrates \cite{andreotti2020statics}. 
Consequently, hydrogels can exhibit time-dependent deformation and flow behavior, which can affect the contact angle and the contact line morphology \cite{chakrabarti2018elastowetting,joanny2001gels}. 
Additionally, the surface chemistry of hydrogels can play a significant role in their interactions with surfaces, 
influencing the strength of adhesion and the wetting behavior with the underlying substrate \cite{yang2020hydrogel,chakrabarti2018elastowetting}.
To disentangle the complex physics involved, 
understanding the force transmission
between a hydrogel and rigid substrate at the three phase contact line is of utmost importance. 
Marchand et al. have shown that for liquid droplets on an elastic substrate, how the force transmission by the liquid on the substrate dictates contact line morphology as well as the contact angles of the different phases involved \cite{marchand2012contact}. 
To date, however, the influence of the droplet elastic modulus $E$ on dictating the force transmission and thus the contact line morphology remains elusive. 
In this Letter, we derive the forces at the contact line between a linearly elastic hydrogel drop and a solid glass substrate using density functional theory (DFT). 
We compare our numerical findings with experimentally obtained static profiles of Polyacrylamide (PAAm) hydrogels of varying elasticity on rigid glass substrates.

\section{II. Experiments}
\begin{figure}
\begin{center}
\includegraphics[width=0.48\textwidth]{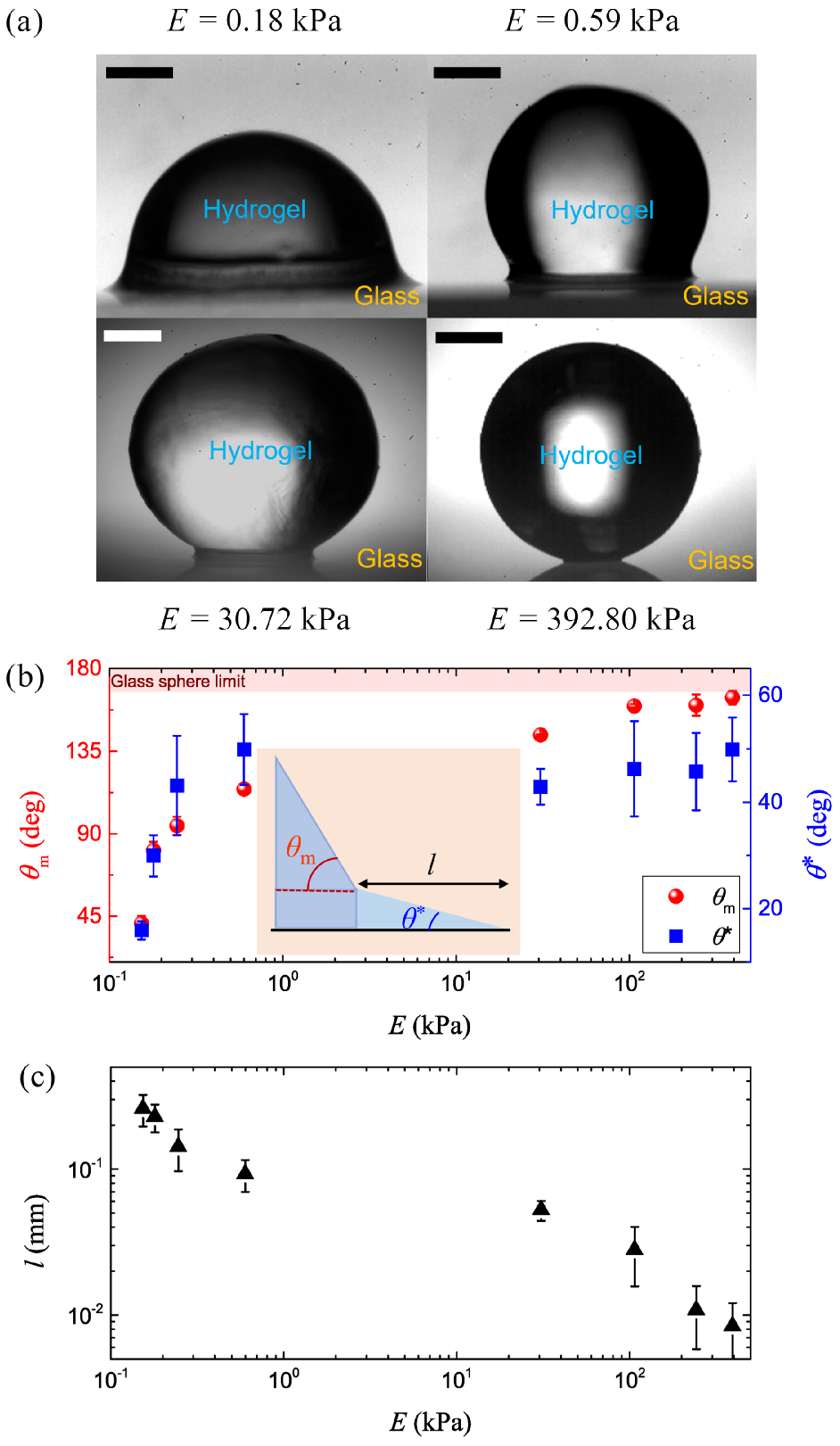}
\caption{(a) Equilibrium snapshots of 1\,mm radius hydrogels of different elastic modulus $E$ on rigid glass substrates. All scale bars represent 0.5\,mm.
(b) Variation of macroscopic contact angle $\theta_{\rm m}$ (left axis) and microscopic foot contact angle $\theta^{*}$ (right axis) with hydrogel elasticity $E$. Inset shows the schematic near of the hydrogel shape near the contact line.
$\ell$ is the foot length. (c) Variation of foot length $\ell$ with hydrogel elasticity $E$. }
\label{fig:1}
\end{center}
\end{figure}

We prepare 1\,mm radius hydrogel drops using acrylamide (AAm) as a monomer, N, N’-Methylene-bis-acrylamide (BIS) as the crosslinker, and 2,4,6-tri-methyl benzoyl-diphenylphosphine oxide (TPO) as the initiator. 
By varying the monomer in weight percentages of 6.3 – 30.0,
we obtain hydrogels with elasticity $E$ varying between 0.15\,kPa to 392.80\,kPa whereas the surface tension varies between 65.7\,mN/m to 57.3\,mN/m (see {\bf{ Appendix A, B}}). 
The static configuration of hydrogel drops on freshly cleaned glass slides was captured using shadowgraphy and subsequently analyzed to extract contact angle information (see {\bf{ Appendix C}}). 
As shown in Fig.~\ref{fig:1}a, the sessile hydrogel drops exhibit decreasing wetting behavior, i.e., increasing contact angle with increasing elasticity $E$. 
For high $E$, the hydrogel configuration resembles that of a rigid glass sphere (Fig.~\ref{fig:1}a, $E = 392.80\,{\rm kPa}$).
 
A distinct feature of the wetting configuration is the microscopic \emph{foot} region which progressively diminishes with increasing hydrogel elasticity.
A natural consequence of the hydrogel \emph{foot} is that the
contact region exhibits a microscopic foot contact angle $\theta^{*}$ which changes curvature away from the contact line into a macroscopic contact angle $\theta_{\rm m}$ (inset of Fig.~\ref{fig:1}b).
We experimentally observe that with increasing hydrogel elasticity, $\theta_{\rm m}$ increases
and approaches the close to 180$^{\circ}$ value normally exhibited by rigid glass
spheres on glass substrates. The microscopic foot contact angle $\theta^{*}$ remains more or less constant for stiffer hydrogels but progressively decreases with decreasing $E$ and converges to the magnitude exhibited by the macroscopic contact angle (Fig.~\ref{fig:1}b). 
Lastly, as shown in Fig.~\ref{fig:1}c, the foot length $\ell$ decreases
with increasing $E$ and may cease to occur in the rigid limit ($E \rightarrow \infty$). On the liquid limit, i.e., $E \rightarrow 0$, the foot also
disappears. In this limit, either the foot merges with the hydrogel bulk or exists in the form of a precursor film. We will elaborate in more detail about the foot features in the two extreme limits later on in the text.

\section{III. Density Functional Theory}
To highlight the effect of hydrogel elasticity on the observed variation of the foot morphology and therein the different contact angles involved, we resort to
the framework of density functional theory (DFT) which takes into account the van der Waals intermolecular forces. 
It should be noted here that we consider the hydrogel as a homogeneous phase throughout our analysis.
Consequently, in a manner consistent with Refs. \cite{getta1998line,marchand2012contact,das2011elastic},
we separate the long-range attractive potential with the hard-core repulsive potential.
In order to regulate the van der Waals repulsive potential in proximity to the constituent molecules, a characteristic length scale must be established \cite{getta1998line,marchand2012contact,das2011elastic,das2013contribution}. 
This length scale $a$ should also account for the interchain distance between polyacrylamide (PAAm) polymers and a typical length of the chain that aligns with the glass-hydrogel interface. Consequently, the interaction potential
between any two phases (1,2) can be expressed as,
\begin{equation}
 {\Phi_{12}}({\rm {r}}) = \rho_{1}\rho_{2}\int_{1}{\rm {dr'}}\tilde{\phi}_{12}|\rm{r-r'}|,
 \label{eqn:1}
\end{equation}
where $\tilde{\phi}_{12}$ is the effective interaction potential between the two phases and $\rho_{1}$, $\rho_{2}$ are their densities. 
The phases can be either of solid, liquid or gas. 
In order to define $\tilde{\phi}_{12}$, we identify an element in phase 1 which is at a distance $\rm{r}$ from phase 2 and $\rm{r'}$ from rest of the constituting elements of phase 1 (see {\bf{ Appendix E}}).
The surface tension can thus be calculated using the effective liquid-liquid
attractive potential ($\tilde{\phi}_{LL}$) as
\begin{equation}
\int_0^{\infty} r^2\,\tilde{\phi}_{LL}(r)\,dr = -\gamma,
\label{eqn:2}
\end{equation}
where the notion of infinity implies large distance away from the contact line relative to the length scale $a$. 
Here, we note that for the ease of convenience we refer to the hydrogel as
the liquid ($L$) phase although strictly speaking, hydrogels exhibit both liquid and solid like properties.

Since $\gamma$ is weakly dependent on elasticity $E$ of the hydrogel (see {\bf{ Appendix B}}), 
equilibrium wetting entails one of the two following events in order to conserve hydrogel volume: (1) At low $E$, the liquid-vapor bulk interface recedes towards the solid which pushes greater volume to the wedge-like foot region. 
Consistent with Eq.~\ref{eqn:2}, $\tilde{\phi}_{LL}$ reduces with decrease in $E$ which in turn relaxes the solid-liquid repulsion within the framework of DFT. Accordingly, a force due to liquid on solid points towards the hydrogel volume. (2) As $E$ increases, reduction in the volume of hydrogel at the contact wedge raises $\tilde{\phi}_{LL}$ which ultimately causes the liquid to repel the solid at the contact. The balance of these forces $f_x$ parallel to the flat substrate scales with the foot length $\ell$ as:

\begin{equation}
    \begin{array}{ccc}
        \ell &\sim& \frac{f_x}{E}\,\ln |y|, 
    \end{array}
    \label{eqn:3}
\end{equation}
where $y$ is normal to the interface. At small and large scales, $y$ is bounded by angles $\theta^*$ and $\theta_{\rm m}$, respectively. $\ell$ satisfies the incompressibility and Hooke's stress $\sigma$ in the bulk which accounts for the rapid change in curvature $\kappa$ at the foot:

\begin{equation}
    \begin{array}{ccc}
        \sigma &=& \gamma_{SL}\,\kappa\,\hat{n},
    \end{array}
    \label{eqn:4}
\end{equation}

where $\gamma_{SL}$ is the solid-liquid interfacial tension and $\hat{n}$ is perpendicular to the interface. These predictions agree well with the numerical outcomes of this Letter.  

\begin{figure*}[]
\begin{center}
\includegraphics[width=1\textwidth]{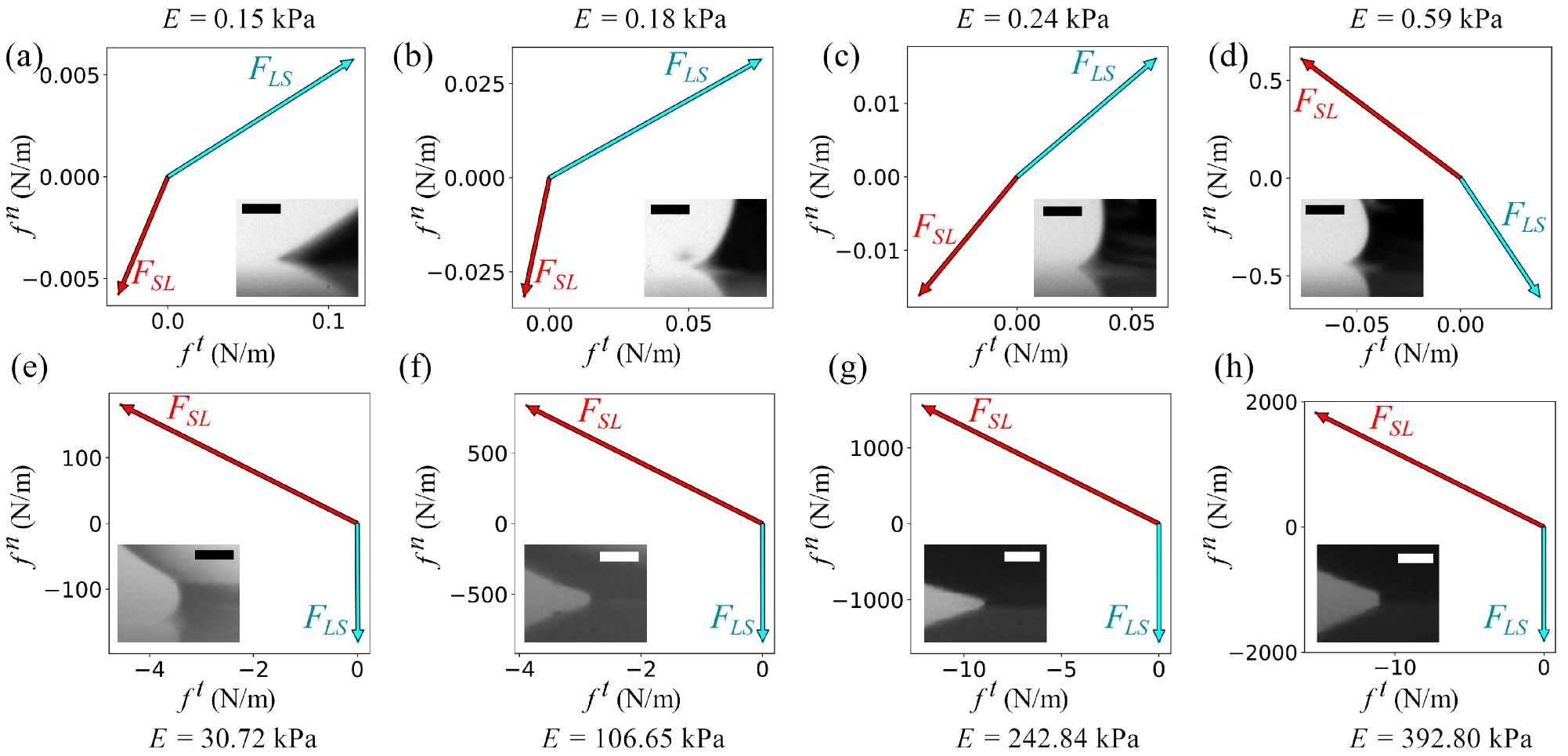}
\caption{Variation of numerically computed normal and tangential force transmission on hydrogels with varying elasticity $E$. Solid-on-liquid and liquid-on-solid forces are represented using $F_{SL}$ and $F_{LS}$, respectively.
Arrows indicate the direction of the net force considering the left edge of contact line. Insets show the experimental snapshots of the hydrogel foot region. Scale bars represent 100\,$\mu$m.}
\label{fig:2}
\end{center}
\end{figure*}
%The contact region where the curvature of the free surface of the sessile hydrogel drop changes rapidly from macroscopic angle $\theta_m$ to contact angle $\theta^*$ is described as the \textquoteleft foot' of the hydrogel. The central result is that there exists a critical elastic modulus $E_c$ below which the forces on the solid and the hydrogel at the contact are attractive in nature, Beyond $E_c$, the forces act repulsively. 

\section{IV. Results and Discussions}
\subsection{A. Forces on the hydrogel and solid at the corner} 
Here, using the liquid-on-solid and solid-on-liquid attractive potentials, i.e.,
$\phi_{LS}$ and $\phi_{SL}$, and the repulsive pressure $p_{r}$, we compute the forces acting on the liquid (hydrogel) wedge due to the solid and vice-versa (see {\bf{ Appendix E}}).
The corner of the liquid wedge experiences a force due to three contributing factors. 
First, a force is exerted by the solid substrate in contact with the hydrogel. Second, an attractive force is exerted by the remaining hydrogel (red region in the inset of Fig.~\ref{fig:1}b). 
Finally, a repulsive force is exerted by the remaining hydrogel due to the flat interface causing an increase in the hydrogel's internal pressure. 
These forces account for the elastic deformations of PAAm hydrogel at the foot of the interfacial contact. 
%Starting from the two-phase attractive potential $\phi$ that defines the surface tension $\gamma$, DFT allows us to quantify the forces that transmit at the corner. 
Consequently, we compute the tangential ($t$) and normal ($n$) components of the
solid-on-hydrogel transmitted force $\overrightarrow{F}_{SL}$ (see Appendix)

\begin{equation}
\ra{1.3}
    \begin{array}{ccc}
         f_{SL}^t &=& \frac{2\,\alpha\,E\,\ell}{3\,\cos \theta^*}  \\
         f_{SL}^n &=& -\frac{\gamma}{\sin \theta^*}(1 - \cos \theta^*\,\cos \theta_{\rm m}) + \frac{4\,\beta\,E\,\ell}{\tan \theta^*}
    \end{array}
    \label{eqn:5}
\end{equation}

such that $F_{SL}=\sqrt{(f_{SL}^t)^2 + (f_{SL}^n)^2}$. 
Here, the constants $\alpha$ and $\beta$ embed the fact that the geometry of the hydrogel foot is a function of $E$. 
Similarly, the components of the hydrogel-on-solid transmitted force $\overrightarrow{F}_{LS}$ are (see {\bf{ Appendix F}})
\begin{equation}
\ra{1.3}
    \begin{array}{ccc}
         f_{LS}^t &=& \gamma\,(1+\cos \theta_{\rm m})  \\
         f_{LS}^n &=& -f_{SL}^n
    \end{array}
    \label{eqn:6}
\end{equation}
such that $F_{LS}=\sqrt{(f_{LS}^t)^2 + (f_{LS}^n)^2}$. 

In Fig.~\ref{fig:2}, we show the numerically computed solid-on-liquid ($F_{SL}$) and liquid-on-solid ($F_{LS}$) forces for hydrogels with different elasticity.
An important outcome follows the computations. 
When the hydrogel has low $E$, the contact forces are attractive in nature which matches with the reported observations using droplet-on-soft solids systems \cite{das2011elastic, marchand2012capillary, marchand2012contact}. 
Interestingly, as $E$ increases, beyond a critical elasticity $E_{c}$, the transmitted forces have a repulsive character. In the present setup, 
$E_c$ lies between 0.24 kPa and 0.59 kPa (Figs.~\ref{fig:2}c-d).
This is the central result of this Letter. 
Evident from Eq.~\ref{eqn:6} and Fig.~\ref{fig:2}, the normal components
of the transmitted forces show attractive character for $E < E_{c}$,
vanishes at $E_{c}$ and switches to being repulsive in nature for  $E > E_{c}$. 
Physically, one may surmise that at low $E$, the hydrogels are liquid-like, and
thus are attracted by the underlying solid. With gradually increasing $E$,
the hydrogels become solid-like and hence exhibit solid-solid repulsion. 

In the liquid limit, $E \rightarrow 0$ and $\theta^{*} \rightarrow \theta_{\rm m}$, 
and thus we have $f_{SL}^{n} \approx -\gamma\sin\theta_{\rm m}$, $f_{LS}^{n} \approx \gamma\sin\theta_{\rm m}$: the vertical force balance of the Young's law \cite{young1805iii,marchand2012contact,marchand2011surface}. 
It is to be noted here that from our experiments using
hydrogels of ultra low elasticity (see Appendix), we observe that the \emph{foot} vanishes in the low elastic limit resembling the perfect spherical cap configuration of liquid drops. Consequently, two possible scenarios can occur in this limit $E \rightarrow 0$: either
the foot merges with the hydrogel bulk providing $\theta^{*} \rightarrow \theta_{\rm m}$ or the foot still exists in the form of a precursor film of nanometric thickness, giving $\theta^{*} \approx 0$. From Eq.~\ref{eqn:5}, we observe that the latter condition presents a singularity and thus is inconsistent with Young's force balance. 
In the rigid limit, i.e., $E \rightarrow \infty$, a singularity arises as well for the condition of no foot, i.e., $\ell \rightarrow 0$ and $\theta^{*} \rightarrow 0$. 
Consequently, we hypothesize that a foot still exists in the rigid limit where its length approaches the molecular length scale, i.e.,  $\ell \rightarrow a$. Further, from our experimental
observation of foot angle saturation for high $E$ (Fig.~\ref{fig:1}b), we can assume that in the rigid limit, the foot poses a finite contact angle, $\theta^{*} \rightarrow 45^{\circ}$.
Thus, in the rigid limit, we can express $E\ell \rightarrow Ea \rightarrow 1$ and $\theta_{\rm m} \rightarrow 180^{\circ}$. 
Therefore, from Eq.~\ref{eqn:5}, $f_{SL}^{n} \approx -2.4\gamma +4\beta \approx -2\gamma$ since $\beta$ is small and consequently, $f_{LS}^{n} \approx 2\gamma$. In other words, the rigid limit implies superior intermolecular packing at the contact which has two important connotations: (1) the overlapping electron clouds approach the repulsive core $a$ and bound the foot length $\ell$. This cut-off is realized parallel to the solid-liquid interface, and not normal to it, because the polymer chains in the hydrogel usually align with the interface and accelerate the intermolecular overlap; (2) dense packing increases the pair correlation function which is implicit in the definition of $\gamma$. Lastly, this finding is consistent with the thermodynamic work of adhesion between any two similar surfaces, i.e., $w = 2\gamma$ \cite{israelachvili2011intermolecular}.

%There exists a critical modulus of elasticity $E_c$ of the hydrogel for which the normal components of the transmitting forces vanish. 
%Physically, one may surmise that hydrogen bonding dominates the low-$E$ regime, while the interactions between the amide group of PAAm hydrogel and the silicon-oxygen group of the glass substrate dictate the force transmission at high-$E$ regime.

Here, we further elaborate on the behavior of the observed forces
with hydrogel elasticity.
From Eq.~\ref{eqn:6} and Fig.~\ref{fig:2}, we observe that for liquid-like softer hydrogels, net liquid-on-solid transmitted force $F_{LS}$ acts towards the interior of the hydrogel drop with
a significant tangential component $f_{LS}^{t}$ at par with existing literature \cite{marchand2012contact,marchand2012capillary,marchand2011surface,das2011elastic}.
With increasing $E$, the tangential component diminishes and approaches zero (since $\theta_{\rm m} \rightarrow 180^{\circ}$) in the rigid limit. 
The observed trend of $f_{LS}^{t}$ can be interpreted in terms of shear: liquids exhibit finite shear whereas solids do not.
On the other hand, the solid-on-liquid tangential force $f_{SL}^{t}$ vanishes 
in the liquid limit ($E \rightarrow 0$). This is reasonable due to the symmetry of the solid substrate on the either side of a sessile liquid drop \cite{marchand2012contact}.

\subsection{B. Hydrogel contact angles}
\begin{figure}
\begin{center}
\includegraphics[width=0.48\textwidth]{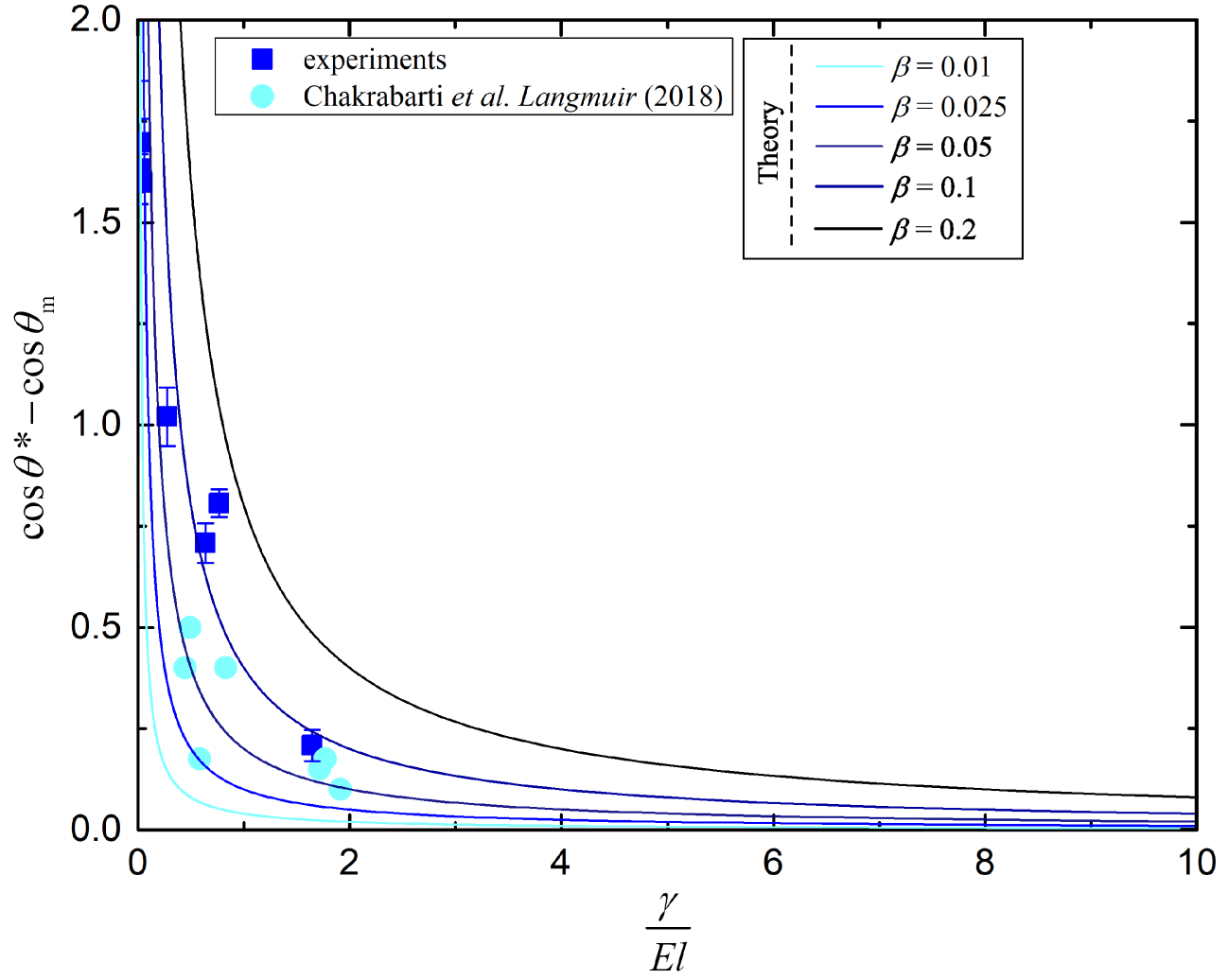}
\caption{Variation of $\cos\theta^{*} - \cos\theta_{\rm m}$ with dimensionless
elastocapillary paramerte $\gamma/{E\ell}$ for hydrogels with different elasticity $E$ and corresponding foot length $\ell$. Literature data from Ref. \citep{chakrabarti2018elastowetting} are also shown. The solid lines represent the theory (Eq.~\ref{eqn:8} in the main text) for different $\beta$ values. }
\label{fig:3}
\end{center}
\end{figure}

Lastly, we recall that the derived forces associate with the hydrogel foot which allows us to identify a wedge-like neighborhood of the foot that is in equilibrium due to the balance of the forces. Since the contact angle $\theta^{*}$ and the macroscopic angle $\theta_{\rm m}$ are associated with the foot as well, the equilibrium provides a unique relation between the two of them.
Here, we first highlight the relations
for the individual contact angles (see {\bf{ Appendix G}} for detailed derivation):
\begin{equation}
\ra{1.3}
    \begin{array}{ccc}
         \cos \theta^* &=& \frac{\alpha}{6\,\beta} \\
         \cos \theta_{\rm m} &=& \frac{\alpha}{6\,\beta} - \frac{4\,E\,\ell\,\beta}{\gamma}
    \end{array}
    \label{eqn:7}
\end{equation}
Interestingly, Eq.~\ref{eqn:7} indicates that the foot angle $\theta^{*}$ is independent of hydrogel elasticity $E$ which is close to our experimental observation where $\theta^{*}$ exhibits a weak dependence on $E$ for majority of the experimental range (Fig.~\ref{fig:1}b).
Subsequently, we can express the relation between $\theta^{*}$ and $\theta_{\rm m}$ as,

%We find that the theoretical behavior of $\theta^*$ and $\theta_m$ as functions of the elastocapillary length scale $\gamma/E$ agree well with our experiments and reported results \cite{chakrabarti2018elastowetting} which, in turn, validate our numerically computed forces at the contact. 
\begin{equation}
   \cos \theta^{*} - \cos \theta_{\rm m} = 4\,\beta\,\frac{\ell}{\gamma/E}.
     \label{eqn:8}
\end{equation}
It is evident from Eq.~\ref{eqn:8} that in the liquid limit ($E \rightarrow 0$), $\theta^{*}$ approaches $\theta_{\rm m}$ which in turn validates our previous assumption. As shown in Fig.~\ref{fig:3}, $\theta^{*}$ asymptotically approaches $\theta_{\rm m}$ with decreasing $E$ (increasing $\gamma/{E\ell}$). 
It is to be noted here that a single $\beta$ value does not satisfy our experimental findings for the entire range of experimental parameters and thus we have used various $\beta$ values to fit our data. However, the range of $\beta$ values used here is consistent with those used for our numerical analysis.
On the other hand, in the rigid limit, where $E \rightarrow \infty$, $E\ell \rightarrow Ea \rightarrow 1$, and $\theta_{\rm m} \rightarrow 180^{\circ}$, we have $\theta^{*} \approx \cos^{-1}(4\beta{\gamma}^{-1} - 1)$. Using $\beta_{\rm avg} \approx 0.125$ (see Appendix) and $\gamma \approx \gamma_{\rm glass} \approx 0.3$, we calculate $\theta^{*} \approx 52^{\circ}$, which is close to our previous assumption of $45^{\circ}$. 
Lastly, we highlight that Eq.~\ref{eqn:8} has interesting implications:
since the L.H.S of Eq.~\ref{eqn:8} is bounded between 0 and 2, the maximum value of $4\beta{{\ell}E/\gamma}$  is 2. Consequently, assuming $\beta_{\rm max} \approx 0.5$ (see Appendix), the maximum value of \emph{foot} length is ${\ell}_{\rm max} \approx \gamma/E$, i.e., the upper bound of hydrogel foot is dictated by the elastocapillary  length-scale.

\section{V. Conclusion}
In conclusion, using theory and experiments, we have highlighted the wetting morphology of elastic hydrogels on rigid substrates. 
Our primary finding indicates a switch of interaction potential between an elastic body and rigid surface
from attractive to repulsive at a critical value of elasticity. 
The finding is fundamentally intuitive as well as crucial towards effective design 
of smart materials for bioprinting \cite{murphy20143d}, contact lenses \cite{ku2020smart}, and soft robotics \cite{whitesides2018soft}. 

\section{VI. Acknowledgments}
The authors thank Haoyuan Jing for fruitful discussions.
P.C., S.M., and S.K.M. acknowledges financial support
from NSERC Discovery (RGPIN-2019-04060).
A.-R.K. is supported by Nanofellowship 2021 of Waterloo Institute for Nanotechnology, University of Waterloo.
B.Z. acknowledges support from NSERC (RGPIN-2019-04650 and RGPAS-2019-00115).

\appendix

\section{Appendix A: Fabrication of hydrogels}
To synthesize hydrogels, the monomer used was acrylamide (AAm).
As the crosslinker, we used N,N’-Methylene-bis-acrylamide (BIS) and as the initiator, we used 2,4,6-tri-methyl benzoyl-diphenylphosphine oxide (TPO) nanoparticle. 
TPO nanoparticles were synthesized by dissolving 2.5 wt.\,\% of diphenyl (2,4,6-trimethylbenzoyl) phosphine oxide 
($\rm {M_{w}} = 348.48$), 3.75 wt.\,\% of polyvinylpyrrolidone, and 3.75 wt.\,\% of dodecyl surface sodium salt (SDS) in  DI water. The ingredients were mixed in a sonicator for 5 min at 95$^{\circ}$C to obtain 10 wt.\,\% of TPO nanoparticles in DI water. 
Pregel solutions were prepared by diluting the monomer (6.3 – 30.0 wt.\%), 1 wt.\% of the crosslinker (based on the monomer), and 2.5 wt. \% of TPO nanoparticles (based on the monomer) in 0.5\,mM NaIO$_{\rm {4}}$ solution (oxygen scavenger). 
We prepared the hydrogel beads by suspending 4\,$\mu$L of the pregel solution in a beaker with n-octane and a silicone oil (phenylmethylsiloxane-dimethylsiloxane copolymer, 500\,cSt, Gelest), with a 1:2 volumetric ratio between the two. 
Spherical shapes of the hydrogel were achieved due to the density gradient between n-octane (density $\rho$\,=\,0.71 g/$\rm {cm^{3}}$) and the silicone oil ($\rho$\,=\,1.08 g/$\rm {cm^{3}}$). 
Each pregel solution was exposed to UV light ($\sim$\,365 nm) for 20 min. 
The cured hydrogel beads were washed with heptane multiple times before each use.
\section{Appendix B: Characterization of hydrogels}
\begin{figure*}[]
\begin{center}
\includegraphics[width=1\textwidth]{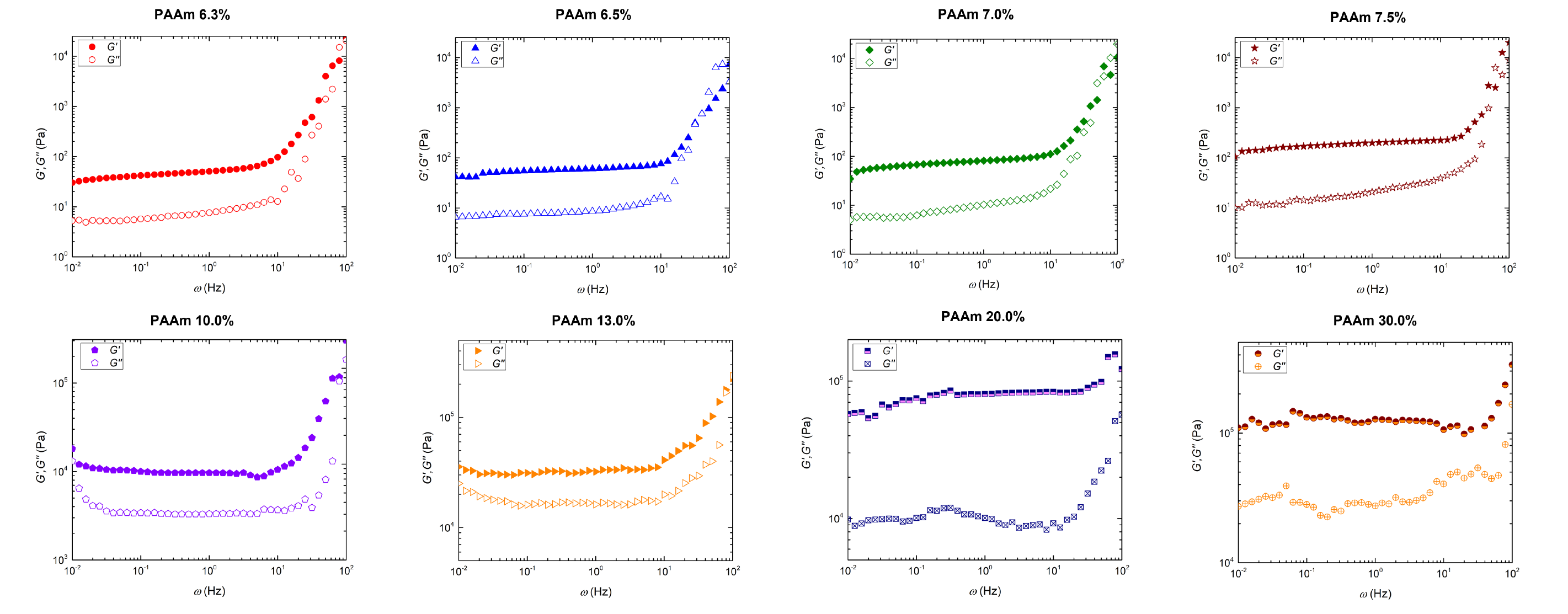}
\caption{Rheology of hydrogels. Variation of storage modulus ($G'$) and loss modulus ($G''$) with frequency $\omega$ for PAAm 6.3 - 30.0\,wt.\% hydrogels.}
\label{fig:S1}
\end{center}
\end{figure*}

To characterize the elasticity of the hydrogels, we performed rheology tests.
For rheology measurements, each PAAm hydrogel was polymerized in a 60 mm-diameter-petri dish with a thickness of 2 mm. 
The cured hydrogels were cut into 25 mm-diameter. 
The shear storage and loss modulus of the materials were measured by performing a frequency sweep test on a dynamic shear rheometer (AR 2000, TA Instruments) from 0.01 to 100 Hz at a strain rate of 1\,\% and a normal force of 1\,N (Fig.~\ref{fig:S1}). 
%({\bf{Supp. Figs. S4-S6}}). 
A constant temperature was maintained at 25$^{\circ}$C, 
and the test adopter is a 25 mm diameter plate. 
The measurement is taken after waiting for 10 min to stabilize the polymer. 
%Each measurement was repeated three times. 

The surface tension of each prepolymer solution was measured using the pendant drop method under ambient condition of 25$^{\circ}$C on a drop-shape analyzer (Kruss, DSA30). 
The surface tension was measured using the Kruss ADVANCE software's in-built Young-Laplace equation (Table.~\ref{tb:hydrogel}).
Each measurement was repeated three times.

%({\bf{Supp. Table. S1}}). 

\begin{table}
	\caption{Physical properties of Acrylamide (AAm) solvent for different monomer weight percentage.
	}
	\centering
	\begin{tabular}{c c c}%\toprule
\hline
\hline
	 {Monomer weight}  & {Surface tension} & Density\\
	 ($\%$) & $\gamma\,({\rm{mN/m}})$  & $\rho\,({\rm kg/m^{3}})$\\
%	 \multirow{2}{*}{Weight ratio}  & \multicolumn{4}{c}{Rigidity modulus $(\omega = 0.01\,{\rm{rad.s^{-1}}})$}  \\
%		                        & $G'\,({\rm{kPa}})$    & $G''\,({\rm{kPa}})$ & $G\,({\rm{kPa}})$ & $E\,({\rm{kPa}})$\\ 		                        
%\bottomrule  \\  
\hline
\hline                    
		  
2.5  & 70.9$\pm$0.2 & 999.7\\[1ex]
4.0  & 69.6$\pm$0.1 & 1001.0\\[1ex]
6.3  & 65.7$\pm$0.1 & 1003.1 \\[1ex]
6.5  & 65.6$\pm$0.2 & 1003.3 \\[1ex]
7.0  & 65.6$\pm$0.1 & 1003.8 \\[1ex]
7.5  & 64.8$\pm$0.3 & 1004.3 \\[1ex]
10.0  & 64.2$\pm$0.2 & 1006.8 \\[1ex]
13.0  & 61.9$\pm$0.1 & 1009.8 \\[1ex]
20.0  & 60.6$\pm$0.2 & 1013.3 \\[1ex]
30.0  & 57.3$\pm$0.1 & 1019.4\\[1ex]
		               
\hline
		
		%\bottomrule
	\end{tabular}
	\label{tb:hydrogel}
\end{table}

\section{Appendix C: Static measurements}
\begin{figure}[]
\begin{center}
\includegraphics[width=.5\textwidth]{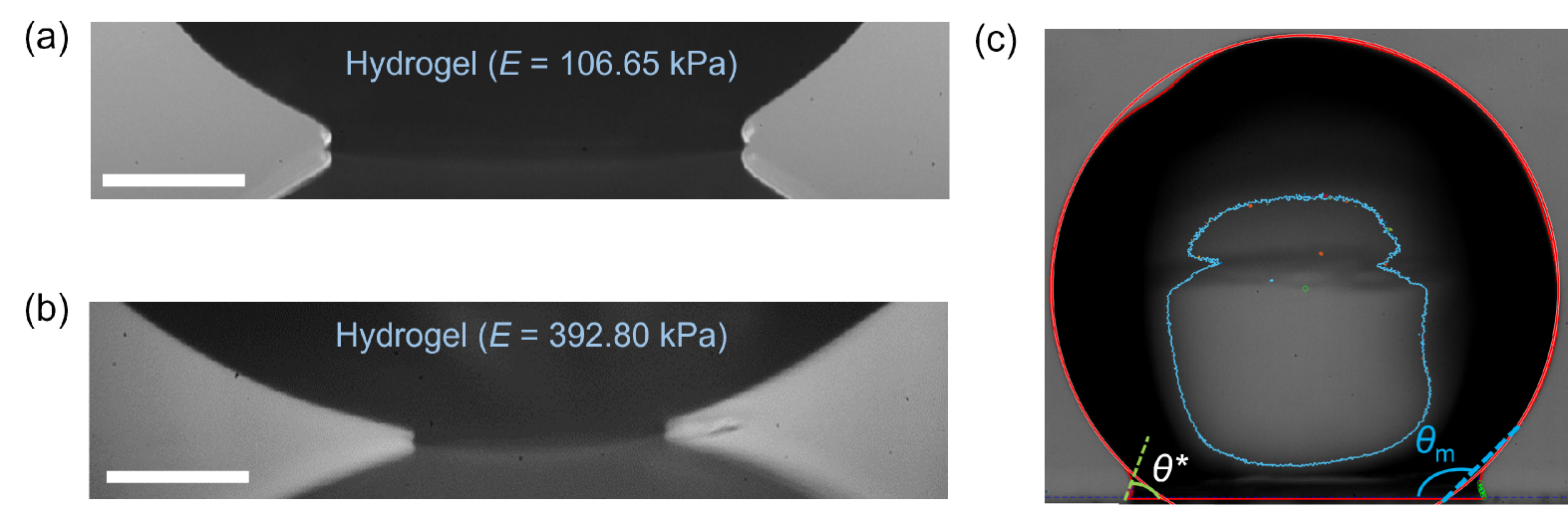}
\caption{(a) Static snapshot obtained using 10x magnificaction of a hydrogel with elastic modulus $E = 106.65\,{\rm kPa}$ on a rigid glass substrate. (b) Static snapshot obtained using 10x magnificaction of a hydrogel with elastic modulus $E = 392.80\,{\rm kPa}$ on a rigid glass substrate. Scale bars represent 250\,$\mu$m. 
(c) Fitting procedure of hydrogel profiles to obtain contact angles.}
\label{fig:S3}
\end{center}
\end{figure}

1\,mm radius hydrogels were deposited on the freshly cleaned glass substrates and made to reach equilibrium. The side view of the equilibrium configuration was imaged using FASTCAM Mini AX high-speed camera (Photron) coupled to a 4x objective lens (Olympus) providing a spatial resolution of 3 – 4\,$\mu$m/pixel. 
Additionally, a higher magnification of 10x (Optem Inc.)
was used for stiffer hydrogels where the foot dimensions are small (Fig.~\ref{fig:S3}a,b).
The equilibrium snapshots were subsequently processed using custom MATLAB routines and ImageJ. 
The contact line was determined from the gray scale intensity profile at the region it presented maximum slope. 
The foot length was determined from the radial location where the liquid-air interface changes curvature.
Alternatively, using the detected boundary of the hydrogel profile, the maximum circle was be fitted, and the corresponding wetting foot and macroscopic contact angle were measured (Fig.~\ref{fig:S3}c).
The foot angle was measured using tangent fit. 
It should be noted here that the dehydration time scale of
the hydrogels are order of magnitude larger than the experimental
time scale and thus does not affect our experimental measurements.

\section{Appendix D: Fabrication and characterization of ultra-low elasticity hydrogels}
\begin{figure}[]
\begin{center}
\includegraphics[width=.5\textwidth]{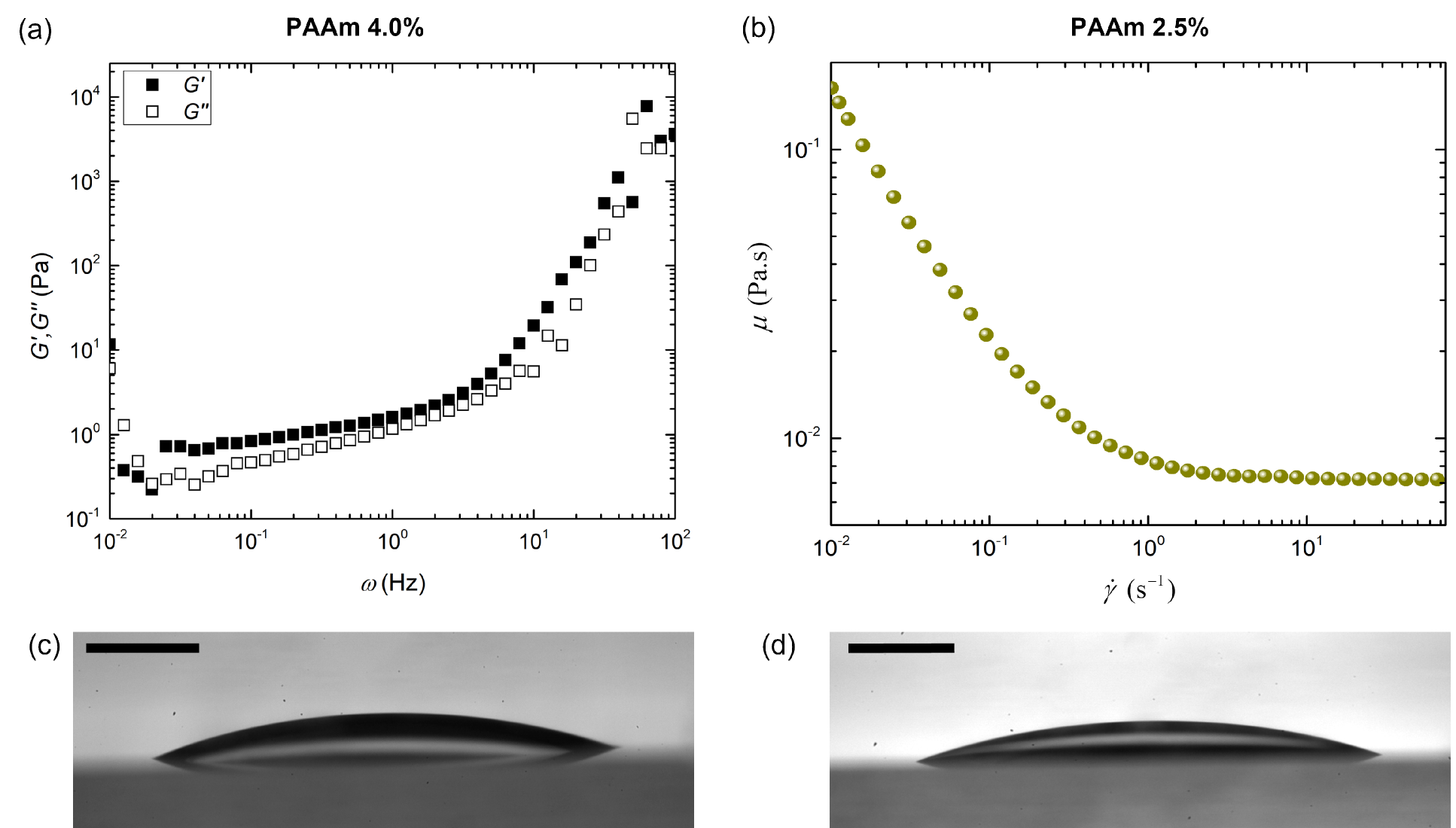}
\caption{(a) Variation of storage modulus ($G'$) and loss modulus ($G''$) with frequency $\omega$ for PAAm 4.0\,wt.\% hydrogel. (b) Variation of shear viscosity $\mu$ with shear rate $\dot{\gamma}$ for PAAm hydrogel with 2.5$\%$ monomer weight percentage. (c)-(d) Static snapshots of PAAm 4.0\,wt.\% and PAAm 2.5\,wt.\% hydrogels. Scale bars represent 1\,mm.}
\label{fig:S2}
\end{center}
\end{figure}

To mimic wetting of liquid drops, we prepared two more hydrogels of
ultra low elasticity using 2.5\% and 4\% monomer wt.\%. 
Rheology measurements of PAAm 4.0 wt. \% indicates an elastic modulus of approximately 5\,Pa (Fig.~\ref{fig:S2}a). 
The cured PAAm 2.5 wt. \% was still liquid, having less than 1\,Pa of the shear storage modulus. As a result, we measured the shear viscosity of PAAm 2.5 wt.\% from 0.01 to 100 Hz using a dynamic shear rheometer (Kinexus Rotational Rheometer, Malvern Instruments) at 25${^\circ}$C and calculated its shear viscosity to be around 0.138\,Pa.s (Fig.~\ref{fig:S2}b). 
Equilibrium configuration of both these hydrogels
exhibited spherical cap like formation with no \emph{foot}, commonly observed for \emph{pure} Newtonian liquid drops (Fig.~\ref{fig:S2}c,d).

\renewcommand{\theequation}{E\arabic{equation}}
\setcounter{equation}{0}

\section{Appendix E: Defining the interaction potential near the contact line}

\begin{figure}[]
\begin{center}
\includegraphics[width=.5\textwidth]{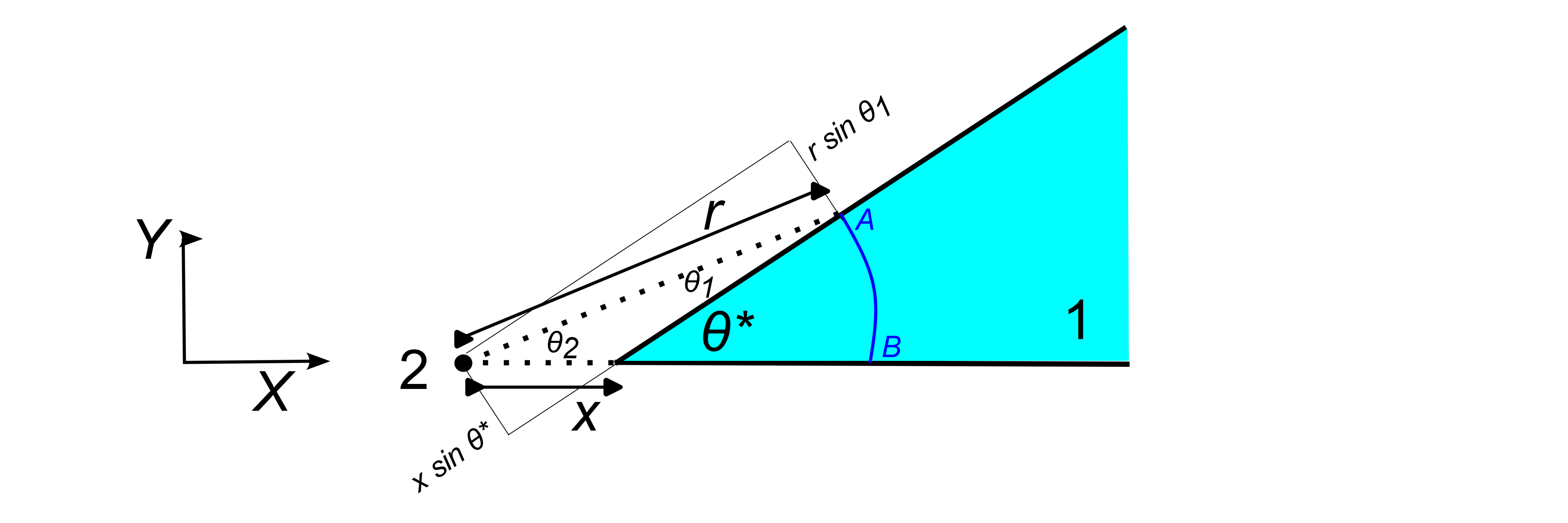}
\caption{Schematic of the wedge to compute the attractive potential.}
\label{fig:supple_1}
\end{center}
\end{figure}

The interaction potential due to phase 1 at a point in phase 2 is the sum of all the infinitesimal potentials due to the extended subsystem of phase 1 that comprises the wedge. As shown in Fig. \ref{fig:supple_1}, the potential due to an element $AB$ of phase 1 at a distance $x$ from the corner of the wedge in phase 2 may be given as:

\begin{equation}
    \begin{array}{ccc}
        d\Phi_{12} (x, \theta^*) &=& r\,\theta_2\,\tilde{\phi}_{12}(r)\,dr
    \end{array}
\end{equation}

where $\tilde{\phi}_{12} = \rho_1\,\rho_2\,\int_{-\infty}^{\infty} \phi_{12}(\sqrt{r^2 + z^2})\,dz$ such that $\phi_{12}$ is the van der Waals interaction potential and z is directed perpendicular to the plane of paper. $\rho_1$ and $\rho_2$ are densities of phases 1 and 2, respectively. From the geometry of exterior-angle property of triangles, $\theta^*=\theta_1 + \theta_2$. In addition, $x\,\sin \theta^* = r\,\sin \theta_1$ which implies that $\theta_2=\theta^* - \sin^{-1}(\frac{x}{r}\,\sin \theta^*)$. Accordingly,

\begin{equation}
    \begin{array}{ccc}
        \Phi_{12} (x, \theta^*) &=& \int_x^{\infty} r[\theta^* - \sin^{-1}(\frac{x}{r}\,\sin \theta^*)]\,\tilde{\phi}_{12}(r)\,dr
    \end{array}
\end{equation}

 Consequently, $\phi_{12}$ and the surface tension $\gamma$ are related by the classical equation \cite{getta1998line,snoeijer2008microscopic}

\begin{equation}
    \begin{array}{ccc}
        \int_0^{\infty} r^2\,\tilde{\phi}_{LL}(r)\,dr &=& -\gamma
    \end{array}
    \label{eq:gamma}
\end{equation}

In Eq. \ref{eq:gamma}, both the phases 1 and 2 correspond to the liquid subsystem.

\renewcommand{\theequation}{F\arabic{equation}}
\setcounter{equation}{0}

\section{Appendix F: Calculation of elastocapillary forces near the contact line}
Forces at the corner of the wedge bear contributions from (1) liquid-vapor, (2) liquid-liquid and (3) solid-liquid interactions. For instance, the force$\overrightarrow{F}_{SL}$ due to solid on liquid may be given as \cite{marchand2012contact}:

\begin{widetext}
\begin{equation}
\begin{array}{ccc}
    \overrightarrow{F}_{SL} &=& -\int_{S_{LV}} \Phi_{SL}^{(1)} (-\sin \theta^* \hat{i} + \cos \theta^* \hat{j}) dS - \\
         \int_{S_{LL}} (\Phi_{SL}^{(2)} - \frac{E}{3\,L}\frac{\partial u}{\partial x}) \hat{i}\, dS + \int_{S_{SL}} (\Phi_{SL}^{(3)} + p_r^{(3)} - \frac{E}{3\,L}\frac{\partial u}{\partial y}) \hat{j}\, dS         
    \end{array}
    \label{eq:F_SL}
\end{equation}
\end{widetext}

\begin{figure*}[]
\begin{center}
\includegraphics[width=0.8\textwidth]{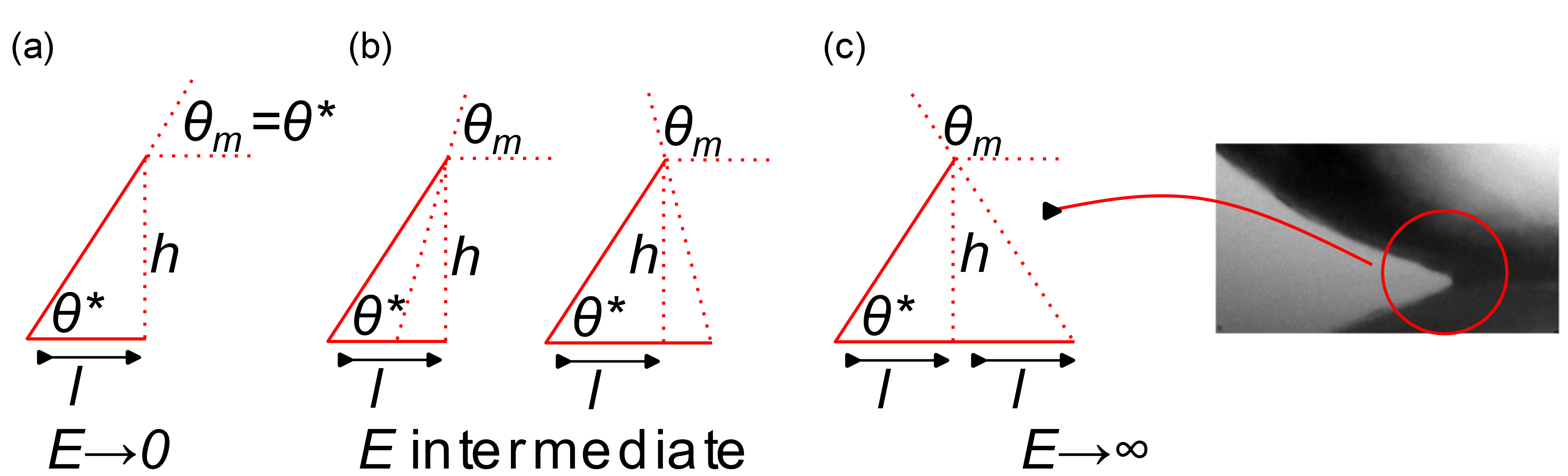}
\caption{Different possible geometry of the foot region and its effect on macroscopic contact angle $\theta_{\rm m}$ depending on hydrogel elasticity $E$. Here, $h$, $l$, and $\theta^{*}$ are the foot height, foot length and foot contact angle, respectively. Inset shows the experimental snapshot of a stiff hydrogel on a glass substrate.}
\label{fig:supple_2}
\end{center}
\end{figure*}

Since the elastic spreading of the hydrogel which produces a foot at the contact should reduce the potential energy of the subsystem, the second and third integrands on the right hand side of Eq. \ref{eq:F_SL} account for the elastic modulation of the interaction potentials. Here, $u$ denotes the displacement of the liquid-vapor interface relative to the section of the fitted circle near the contact line, and $L$ in the denominator denotes the circumference of the solid-liquid interface. Further, let us assume that the boundary of the fitted circle in the absence of the foot, the actual foot boundary and the substrate constitute an isosceles triangle in the glass limit ($E\rightarrow \infty$) as shown in Fig \ref{fig:supple_2}. As the position $x$ increases from zero at the start of the foot till the distance $\ell$ that corresponds to the projection of the topmost vertex of the triangle, the deformation $u$ decreases from $2\,\ell$ near the substrate to zero at the above vertex. Thus, $\frac{\partial u}{\partial x}=-\frac{2\,\ell}{\ell}=-2$. Simultaneously, perpendicular to the substrate, the distance $y$ increases from zero to height $h$ of the foot which implies $\frac{\partial u}{\partial y}=-\frac{2\,\ell}{h}=-\frac{2}{h/\ell}=-\frac{2}{\tan \theta^*}$. At intermediate values of $E$, the geometry of the foot will deviate from being an isosceles triangle, as shown in Fig. \ref{fig:supple_2}, in which case: 

\begin{equation}
\ra{1.3}
    \begin{array}{ccc}
        \frac{\partial u}{\partial x} &=& -2\,\alpha \\
        \frac{\partial u}{\partial y} &=& -\frac{2\,\beta}{\tan \theta^*}
    \end{array}
\end{equation}

where $\alpha$ and $\beta$ are functions of the modulus of elasticity $E$ and account for the shape of the contact foot. 
In our simulations, $\alpha$ and $\beta$ lie in the intervals $[0.01, 1.0]$ and $[0.1, 1.0]$, respectively.

Separating the orthogonal components, we can write $\overrightarrow{F}_{SL} = f_{SL}^t\,\hat{i} + f_{SL}^n\,\hat{j}$, where

\begin{equation}
\ra{1.3}
    \begin{array}{ccc}
        f_{SL}^t &=& -\frac{2\,E\,\ell}{3\,\cos \theta^*}\,\alpha\\
        f_{SL}^n &=& -\frac{\gamma\,(1 - \cos \theta^*\,\cos \theta_{\rm m})}{\sin \theta^*} + \frac{4\,E\,\ell}{\tan \theta^*}\,\beta
    \end{array}
\end{equation}

Similarly, $\overrightarrow{F}_{LS} = f_{LS}^t\,\hat{i} + f_{LS}^n\,\hat{j}$, where

\begin{equation}
\ra{1.3}
    \begin{array}{ccc}
        f_{LS}^t &=& \gamma\,(1 + \cos \theta_{\rm m})\\
        f_{LS}^n &=& - f_{SL}^n
    \end{array}
\end{equation}

\renewcommand{\theequation}{G\arabic{equation}}
\setcounter{equation}{0}

\section{Appendix G: Calculation of contact angle relations}

Since the subsystem of liquid that comprises the contact foot is at equilibrium, we can balance the forces which are parallel and perpendicular to the interface:

\begin{equation}
    \begin{array}{ccc}
        \gamma\,(1 + \cos \theta^*) - \gamma\,(1 + \cos \theta_{\rm m}) + f_{SL}^t &=& 0 \\
        \gamma\,\sin \theta^* + f_{SL}^n &=& 0
    \end{array}
\end{equation}

The force balance ultimately results in governing equations for $\theta^*$ and $\theta_{\rm m}$:

\begin{equation}
\ra{1.3}
    \begin{array}{ccc}
         \cos \theta^* &=& \frac{\alpha}{6\,\beta} \\
         \cos \theta_{\rm m} &=& \frac{\alpha}{6\,\beta} - \frac{4\,E\,\ell\,\beta}{\gamma}
    \end{array}
    \label{eq:theta_star_theta_m}
\end{equation}

\begin{figure}[]
\begin{center}
\includegraphics[width = 0.5\textwidth]{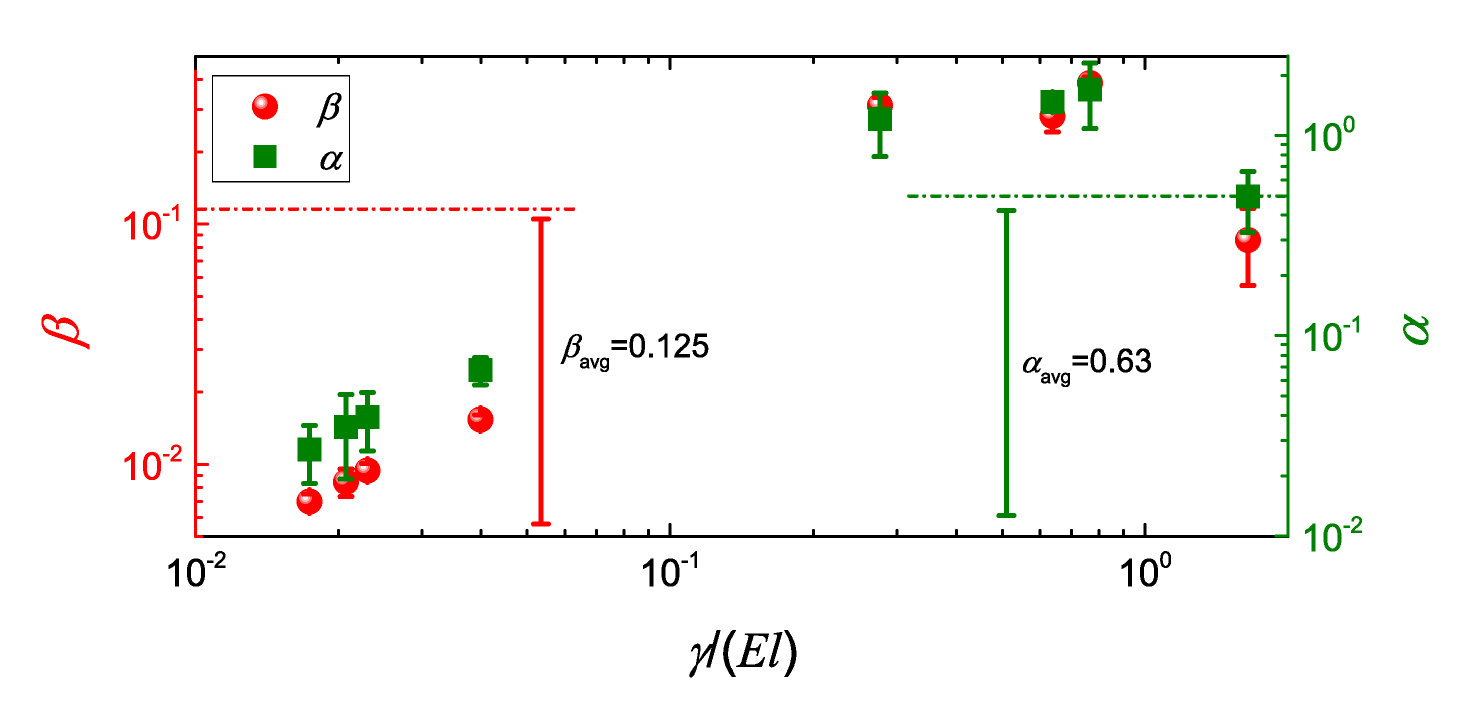}
\caption{Parameters $\alpha$ and $\beta$ as a function of
dimensionless elastocapillary parameter $\gamma/{El}$ calculated using experimental values of $\theta^{*}$, $\theta_{\rm m}$, $\ell$, and $E$ using Eq.~\ref{eq:theta_star_theta_m}.}
\label{fig:supple_3}
\end{center}
\end{figure}

Equation \ref{eq:theta_star_theta_m} may be compacted as $\cos  \theta^{*} - \cos \theta_{\rm m} = \frac{4\,E\,\ell\,\beta}{\gamma}$ that appears in the main text.
Here, we note that upon using Eq.~\ref{eq:theta_star_theta_m} to evaluate $\alpha$ and $\beta$ using our experimentally observed contact angle values, we observe that experimentally calculated values agrees well with the individual bounds used for simulation (Fig.~\ref{fig:supple_3}). 

\bibliography{ref_dft}% Produces the bibliography via BibTeX.

\end{document}